\documentclass[a4paper, 11pt, conference]{ieeeconf}

\IEEEoverridecommandlockouts
\usepackage{geometry}
\usepackage{graphicx}
\usepackage{mathptmx}
\usepackage{times}
\usepackage{amsmath,amssymb}
\usepackage{cite}
\usepackage{url}
\usepackage{etoolbox}
\patchcmd{\keywords}{\textit{Index Terms}}{\textit{Keywords}}{}{}
\begin{document}\raggedbottom
\setlength{\textfloatsep}{12pt plus 2pt minus 2pt}

\title{\LARGE\bfseries
A User-customized and Untethered Electro-haptic Device for Immersive Human-Machine Interaction}
\author{%
  Ziang Cui$^{1,*}$, Siyuan Chen$^{2,*}$, Shanyong Wang$^{3}$, Yining Zhao$^{3}$, Yiran Wang$^{4}$,\\
  Xingming Wen$^{1}$, and Ze Xiong$^{1,\dagger}$%
  \thanks{$^{1}$ School of Biomedical Engineering, ShanghaiTech University,
Shanghai, China.}%
\thanks{$^{2}$ School of Physical Science and Technology, ShanghaiTech University,
Shanghai, China.}%
\thanks{$^{3}$ School of Information Science and Technology, ShanghaiTech University,
Shanghai, China.}%
\thanks{$^{4}$ Electrical Engineering Department, Columbia University,
New York, NY, USA.}%
\thanks{$^{*}$ Ziang Cui and Siyuan Chen contributed equally to this work.
$^{\dagger}$ Ze Xiong is the corresponding author
(\texttt{xiongze@shanghaitech.edu.cn}).}%
  \thanks{This work was supported by National Natural Science Foundation of China (22574106, T2522023), the Science and Technology Commission of Shanghai Municipality (24490710900, 25XF3201800).}%
}

\maketitle
\thispagestyle{empty}

\begin{abstract}
Haptic feedback is essential for human--machine interaction, as it bridges physical and digital
experiences and enables immersive engagement with virtual environments. Current haptic devices
are frequently tethered and lack portability and flexibility. They also have limited
ability to deliver fine-grained, multi-dimensional feedback. To address these challenges, we
present a flexible, ultra-thin, and user-customized electro-haptic device fabricated with soft
materials and printable liquid metal ink. Its highly integrated and lightweight design
minimizes interference with natural hand movements while maintaining reliable skin contact. By
delivering finely controlled electrical stimulation through 15 electrodes, it can evoke a wide
range of tactile sensations that cover diverse interaction scenarios.
\end{abstract}

\begin{keywords}
Electro-haptics, Wearable Devices, Flexible Electronics, Virtual Reality
\end{keywords}

\section{Introduction}

Hand-worn haptic interfaces for virtual reality (VR) should deliver tactile cues across the
palm and fingers while preserving natural hand motion. Previous approaches have addressed
wearability with skin-integrated actuators~\cite{ref61}, thin untethered actuators~\cite{ref23},
and gloves driven by fluidic arrays~\cite{ref50}. Wearable gloves have also combined motion
sensing with haptic feedback for VR~\cite{ref66}. Electro-haptic interfaces use controlled
electrical pulses to evoke tactile sensations~\cite{ref26,ref39}. Prior work has demonstrated
rendering of roughness and other object properties~\cite{ref03,ref60}, thin conformable
electrodes~\cite{ref56}, and multimodal feedback with a wearable glove~\cite{ref28}.

Full-hand electrode arrays and stimulation from the back of the hand have extended tactile
coverage~\cite{ref01,ref53}, while current steering has improved spatial resolution~\cite{ref35}.
Hand dimensions also vary with posture~\cite{ref31}, and individualized geometry has been
used to design flexible tactile-sensing gloves~\cite{ref41}. For an electro-haptic device,
the device contour and electrode layout must fit the user's hand so that stimulation reaches
the intended sites. This requires translating a digital hand model into a manufacturable
contour, electrode positions, and conductive traces.

\begin{figure}[tb]
  \centering
 \includegraphics[width=\columnwidth,height=100mm,keepaspectratio]{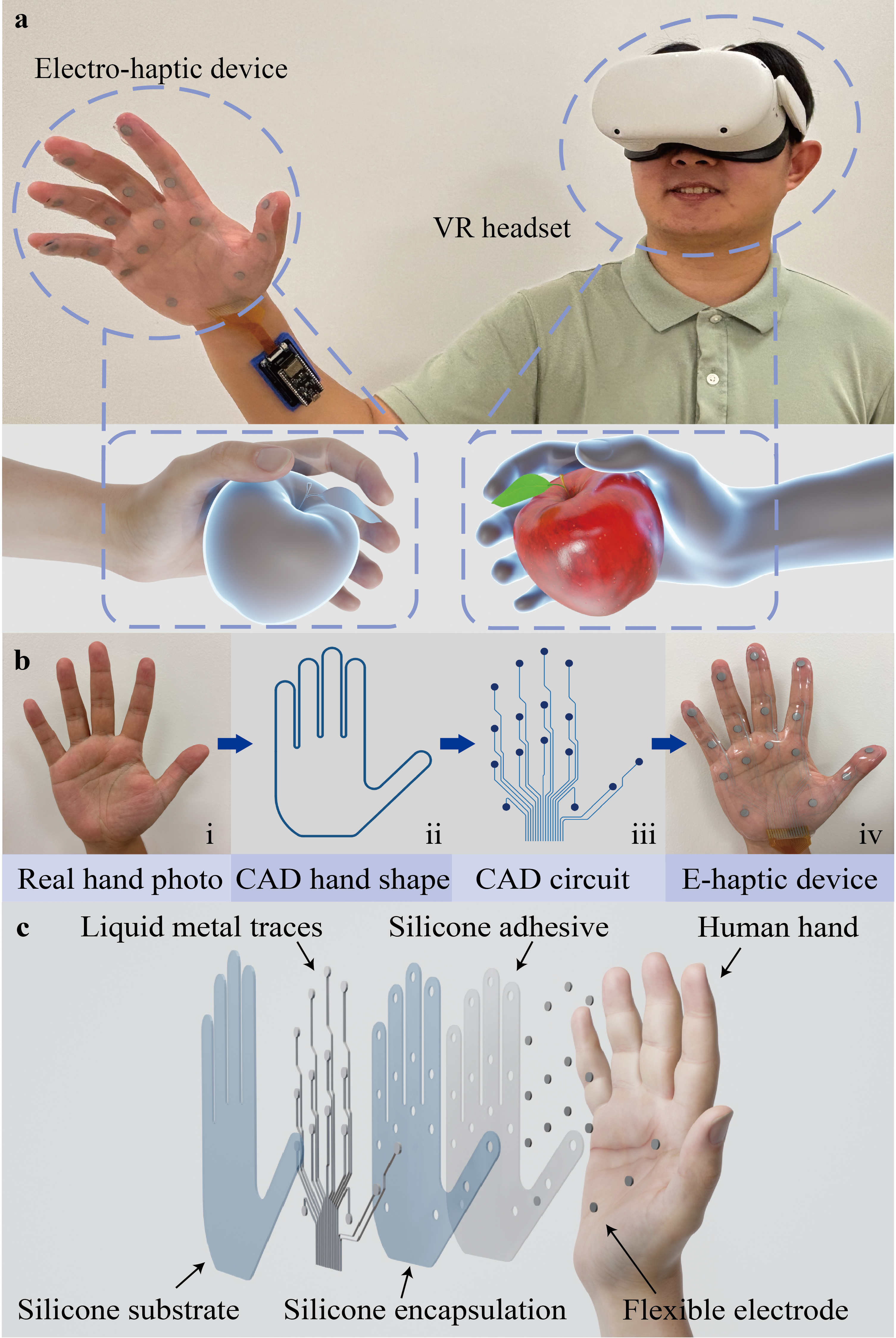}
  \caption{(a) Engaging in immersive VR experience using our electro-haptic device. When interacting with
  objects in the VR environment, even though users are not physically holding anything in the
  real world, the electro-haptic device provides them with a realistic haptic sensation, as if
  they were holding actual objects in their hands. (b) The customized design workflow for our
  electro-haptic device. Simply by capturing an image of the user's hand, an appropriately sized
  model can be generated and used for device fabrication. (c) Schematic diagram illustrating the
  structure of our electro-haptic device.}
  \label{fig:overview}
\end{figure}

In this work, we present an untethered, ultra-thin electro-haptic device with a contour and electrode layout
customized from a single photograph of the user's hand (Fig.~\ref{fig:overview}). Printable
liquid metal--montmorillonite ink forms the conductive traces, and a wrist-worn stimulator
selects from 15 stimulation electrodes distributed across the palm and fingers. The mean
relative error in finger-length estimation was below 2.5\% across the five fingers. We
demonstrate the device in VR grasping, basketball shooting, and painting.

\section{Methods}

\subsection{Personalized Design Algorithm}

Our personalization workflow builds on MobRecon~\cite{ref09} to reconstruct a 3D hand mesh
from a single red--green--blue (RGB) image. From just one photograph, the method produces both a 2D hand contour for accurate glove fitting and a 3D mesh 
compatible with Unity-based customization. We further evaluate its performance to demonstrate effectiveness.

To acquire input data, the users place their right hand flat on a surface, palm facing
up. A camera about 40~cm above the hand captures a well-lit square image. The captured image is then processed by our algorithm, which extracts key features and performs the subsequent steps required for 3D reconstruction. The algorithm comprises several key steps:

\subsubsection{2D Encoding}

We first extract 2D features from the input hand image using a lightweight stacked encoder.
This design reduces parameters while preserving essential geometric details.

\subsubsection{Feature Lifting}

Spatial regression combines heatmap encoding with position regression to improve localization
accuracy and maintain temporal consistency. Semantic transformation converts 2D pose features
into semantically meaningful representations of spatial relationships between hand parts.
Vertex mapping uses a learnable mapping matrix to project 2D features onto 3D mesh vertices
for accurate alignment.

\subsubsection{3D Decoding}

We apply an efficient graph operator to reconstruct a detailed 3D hand mesh from the lifted
features.zW

\subsubsection{Model Fitting}

The reconstructed mesh follows the MANO hand topology~\cite{ref47} and consists of
778 vertices and 1538 faces. To improve accuracy, the reconstructed mesh is
refined by fitting the joint coordinates to a reference plane and projecting the vertices
accordingly, ensuring consistent geometric alignment with the real hand.

\paragraph{Joint Coordinate Regression} Given the 21 joint coordinates (20 hand joints plus one
at the palm root), denoted as $\mathbf{p}=\{p_1,p_2,\ldots,p_N\}$, where each
$p_i\in\mathbb{R}^3$, we fit these points to a plane using a least-squares regression approach.
The plane is defined by $\mathbf{n}\cdot\mathbf{p}+d=0$, where
$\mathbf{n}=(n_x,n_y,n_z)$ is the normal vector of the plane, and $d$ is the distance from the
origin to the plane.

\paragraph{Covariance Matrix and Eigenvalue Decomposition} To determine the optimal plane, we
compute the centroid of the joint coordinates as
\begin{equation}
  \mathbf{c}=\frac{1}{N}\sum_{i=1}^{N}\mathbf{p}_i.
  \label{eq:centroid}
\end{equation}

Next, we construct the covariance matrix $\mathbf{C}\in\mathbb{R}^{3\times 3}$ as
$\sum_{i=1}^{N}(\mathbf{p}_i-\mathbf{c})(\mathbf{p}_i-\mathbf{c})^{T}$. The normal vector
$\mathbf{n}$ of the plane is obtained by performing eigenvalue decomposition on $\mathbf{C}$.
Specifically, $\mathbf{n}$ corresponds to the eigenvector associated with the smallest
eigenvalue, which minimizes the orthogonal distance from the joint points to the plane.

\paragraph{Projection of Mesh Vertices} Once the plane is determined, we project the 778 mesh
vertices $\mathbf{V}=\{\mathbf{v}_1,\mathbf{v}_2,\ldots,\mathbf{v}_{778}\}$ onto the plane. The
projection of a vertex $\mathbf{v}_i$ onto the plane is
\begin{equation}
  \mathbf{v}'_i=\mathbf{v}_i-(\mathbf{n}\ast(\mathbf{v}_i-\mathbf{c}))\mathbf{n},
  \label{eq:vertex-projection}
\end{equation}
where $\mathbf{v}'_i$ is the projected vertex. This ensures that the final 3D model is both
accurate and aligned with the physical hand. By leveraging this regression-based fitting
process, we achieve a robust alignment of the hand model with the physical hand, ensuring that
the 778 vertices are optimally projected onto the fitted plane.

\subsection{Material Design}

To fabricate a flexible and ultra-thin electro-haptic device, we have undertaken an optimized
design from a material perspective. The core component of our approach is a printable flexible
conductive ink. This printable design significantly reduces the manufacturing complexity of our
device and allows for better adaptation to customized device shapes. By combining flexible
encapsulation materials and an extremely low-cost circuit control unit, we have substantially
reduced the cost of the device, as well as its size (particularly thickness), weight, and
fabrication difficulty.

\subsubsection{Design of Printable Liquid Metal Ink}

In previous work, similar electrical stimulation haptic devices have utilized thin metal layers or conductive
coatings~\cite{ref04,ref43}. Gallium-based liquid metals combine metallic conductivity
with the ability to deform~\cite{ref13,ref14}. For eutectic gallium--indium (EGaIn), the
conductivity is approximately $3.4\times10^{6}\,\mathrm{S}/\mathrm{m}$~\cite{ref14}.
The high surface tension of liquid metal makes it difficult to print high-resolution
patterns~\cite{ref15}. Our device uses a printable liquid metal--montmorillonite ink
to address this fabrication challenge~\cite{ref12}.

\begin{figure}[tb]
  \centering
  \includegraphics[width=\columnwidth,height=100mm,keepaspectratio]{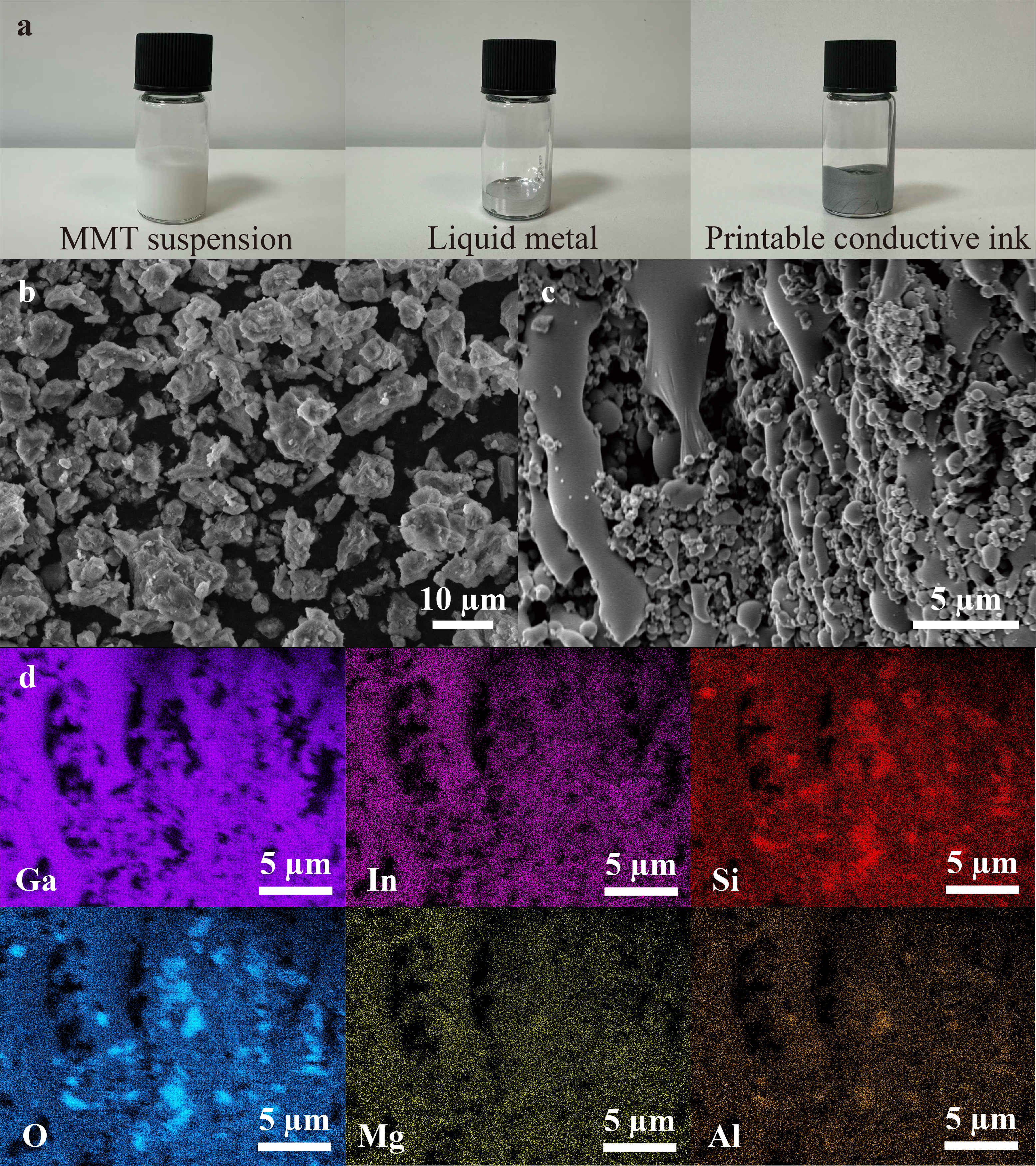}
  \caption{Design of printable liquid metal conductive ink. (a) Montmorillonite dispersion,
  liquid metal, and printable liquid metal--montmorillonite conductive ink. (b) SEM images of montmorillonite at two magnifications. (c) Surface and
  cross-sectional SEM images. (d) EDS images for Ga, In, Si, O,
  Al, and Mg.}
  \label{fig:ink}
\end{figure}

In our work, we employed a montmorillonite dispersion to facilitate the production of liquid
metal conductive ink (Fig.~\ref{fig:ink}). 2~g of montmorillonite was dispersed in 20~g of
deionized water using an ultrasonic disruptor. Subsequently, 10~g of liquid metal (EGaIn) was
added, and the ultrasonic disruptor was used again to process the liquid metal into
micron-sized microspheres. Through centrifugation, excess deionized water and a small amount of
montmorillonite were removed, leaving the precipitated liquid metal-montmorillonite composite
ink at the bottom. The ink was loaded into syringes and mounted on a pneumatic printer. By
controlling the printing speed and extrusion pressure, pre-designed patterns could be printed.
The micrographs and elemental maps in Fig.~\ref{fig:ink}(b)--(d) were acquired
at an accelerating voltage of 5~kV using a scanning electron microscope
(JSM-7800F, JEOL) equipped with an energy-dispersive X-ray spectroscopy (EDS) detector.

In the composite ink, the liquid metal is maintained as
micron-sized microspheres due to the coating of the gallium oxide shell. However, the presence
of gallium oxide limits the conductivity of the composite ink. Patterns printed with this
composite ink are initially almost non-conductive. Heat treatment of the pattern causes the
water content in the ink to evaporate, reducing the pattern's thickness, and allowing the
montmorillonite and liquid metal microspheres to sediment densely. Under the sedimentation
pressure from the montmorillonite, the gallium oxide shells on the liquid metal microspheres
are compressed and rupture, allowing the liquid metal to re-form interconnected conductive
pathways.

Through this design, we maximize the high conductivity and fluidic properties of liquid metal,
processing it into a printable ink. We can pre-design the conductive pathways within the device
and directly print them onto a flexible substrate using a pneumatic printer. Through a
post-processing step, the ink forms well-conducting paths on the substrate.

\subsubsection{Design of Electrode Material}

On-skin electrodes need conformal contact for electrical sensing and stimulation. Rigid
metal electrodes can require conductive gel, which adds a step when putting on the device
and can complicate long-term use~\cite{ref57,ref19}. We selected a flexible electrode
composite to improve conformity with the skin.

Considering that the device encapsulation material uses silicone rubber (Smooth-On, Ecoflex
00-30), we selected a composite material consisting of conductive fillers and silicone rubber
for the electrode. The silicone rubber serves as the flexible outer encapsulation matrix. The
conductive materials were a combination of liquid metal and carbon nanotubes. 2~g of liquid
metal, 1~g of carbon nanotubes, and 10~g of Ecoflex 00-30 Part A were added together and mixed
using a planetary mixer. Subsequently, 10~g of Ecoflex 00-30 Part B was added, and the mixture
was homogenized again using the planetary mixer. The well-mixed slurry was cast into a film and
cured at room temperature for 4 hours. After curing, it can be cut into the required electrode
shapes using a UV laser (LPKF U4) for subsequent use.

\subsection{Wrist-worn Electro-haptic Stimulator}

The wrist-worn electro-haptic stimulator adopts a minimalist design to minimize the
manufacturing cost, volume, and weight of the control circuit as much as possible. It can
perform boost conversion and deliver current output to 15 electrodes. Crucially, it allows for
control of the current magnitude, thereby ensuring the safety of the electro-haptic feedback.

\subsubsection{Circuit Design}

\begin{figure}[tb]
  \centering
 \includegraphics[width=\columnwidth,height=100mm,keepaspectratio]{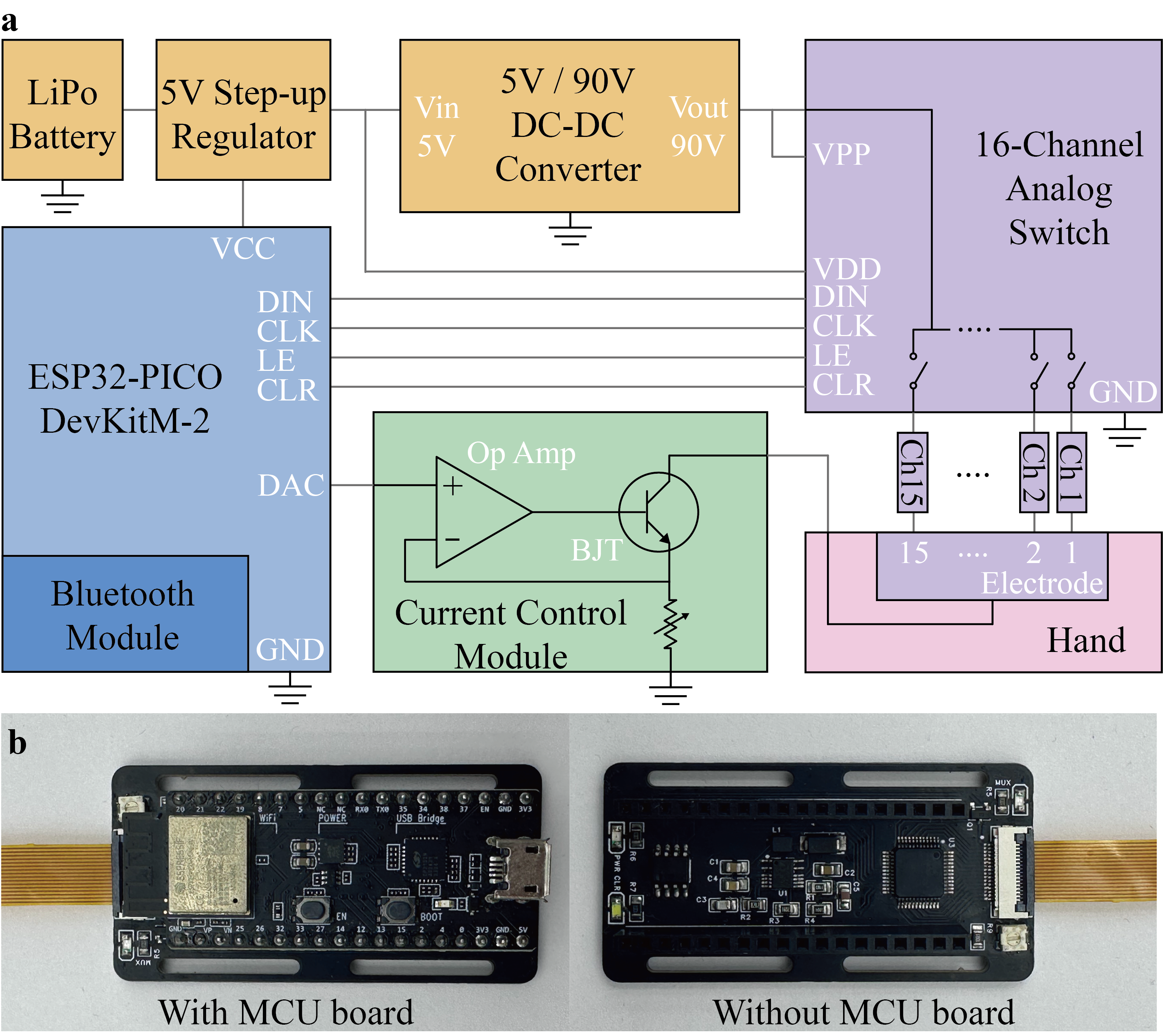}
  \caption{(a) Schematics of our custom electro-haptic stimulator. (b) Our wrist-worn electro-haptic
  stimulator.}
  \label{fig:stimulator}
\end{figure}

The complete dimensions of our hardware circuit are 30~mm $\times$ 60~mm $\times$ 18~mm, with a
weight of only 29~g. As shown in Fig.~\ref{fig:stimulator}, we use a 3.7~V lithium polymer
battery to power the entire system. Through a boost regulator (PW5100, PWChip), the 3.7~V input
voltage is raised to 5~V. The 5~V voltage supplies the next stage DC-DC boost converter
(LT8365, Analog Devices) and the microcontroller (ESP32-PICO-DevKitM-2, Espressif Systems).
After voltage conversion, the 5~V voltage is boosted to 90~V to generate sufficient current for
haptic perception. The microcontroller utilizes an onboard LDO chip to convert the 5~V supply
to the 3.3~V required by the SoC chip. The microcontroller incorporates a Bluetooth module for
wireless communication. The 90~V high voltage is supplied to a MUX chip (TMUX9616, Texas
Instruments). Under the logical control of the microcontroller, the 15 channels are selectively
turned on to generate electrical stimulation haptic sensations on the hand. The current from
the hand returns to the circuit through a virtual ground (VGND) electrode. The current passing
through the hand is controlled by a constant current source circuit composed of an operational
amplifier (LMV358) and a BJT transistor (MMBT5551).

\subsubsection{Generation of Electrical Stimulation Pulses}

When our microcontroller receives serial information from the host computer via the Bluetooth
module, it processes the information and transmits it to the MUX to control the turning on and
off of the 15 channels. Through logical control input from the microcontroller, the period and
duty cycle of each pulse signal can be precisely controlled, enabling us to achieve more
nuanced haptic feedback. Due to the excellent switching performance of the MUX, each channel
can switch states within 1.5 microseconds, allowing us to design different pulse patterns well
within the temporal recognition threshold of haptics. In our design, only one channel is active
at any given time, while the others remain off during this period. This ensures better control
over the haptic intensity and prevents interference between multiple channels.

\subsubsection{Safety Assurance}

To ensure the safe operation of the system, we use a constant current source circuit to control
the magnitude of the current flowing through the hand. The microcontroller's built-in DAC
outputs a control voltage to the non-inverting input of the operational amplifier. The output
of the operational amplifier is connected to the base of the BJT. The inverting input of the
operational amplifier is connected to the emitter of the BJT, and the collector of the BJT is
connected to the VGND electrode on the hand. A variable resistor is connected between the
emitter and ground for manual adjustment of the current level. When the BJT is conducting, the
voltage across the variable resistor remains constant, and the current flowing through the
resistor is limited by its resistance value. Since the current from the hand can only flow
through this loop, the current passing through the hand is likewise constrained to a constant
value. During the experiments, we set the maximum output current to 3~mA.

\subsection{Sequential Device Fabrication}

Based on the device morphology design algorithm mentioned previously, a user’s hand image must first be acquired. The image requirements are detailed in the earlier section. After processing by the algorithm, the joint coordinates of the user’s hand are obtained. These coordinates are used to determine information such as palm width, finger length, and electrode placement, thereby enabling the customized design of the device. Considering potential issues during
fabrication, utilizing a standardized device morphology is essential. The hand was simplified
into a combination of straight lines and arcs, and a computer-aided design (CAD) pattern
was generated from the hand dimensions acquired with the algorithm.

Considering the impact of hand movement on the device positioned on the palm, electrodes were
placed on the fingertips, the areas of the fingers near the palm, and the finger root regions
to avoid separation of the electrodes from the hand caused by motion. Based on the joint
coordinates obtained from the algorithm, a CAD file specifying the electrode positions was
generated. Considering the electrical connections on the device, the connection interface was
uniformly designed near the base of the palm close to the wrist. Conductive trace paths were
designed according to the correspondence between electrodes and this interface. The paths must
remain within the boundaries of the device morphology and avoid crossings to ensure correct
device functionality. The trace paths were also generated in CAD format to facilitate
subsequent printing of the conductors.

\begin{figure}[tb]
  \centering
  \includegraphics[width=\columnwidth,height=100mm,keepaspectratio]{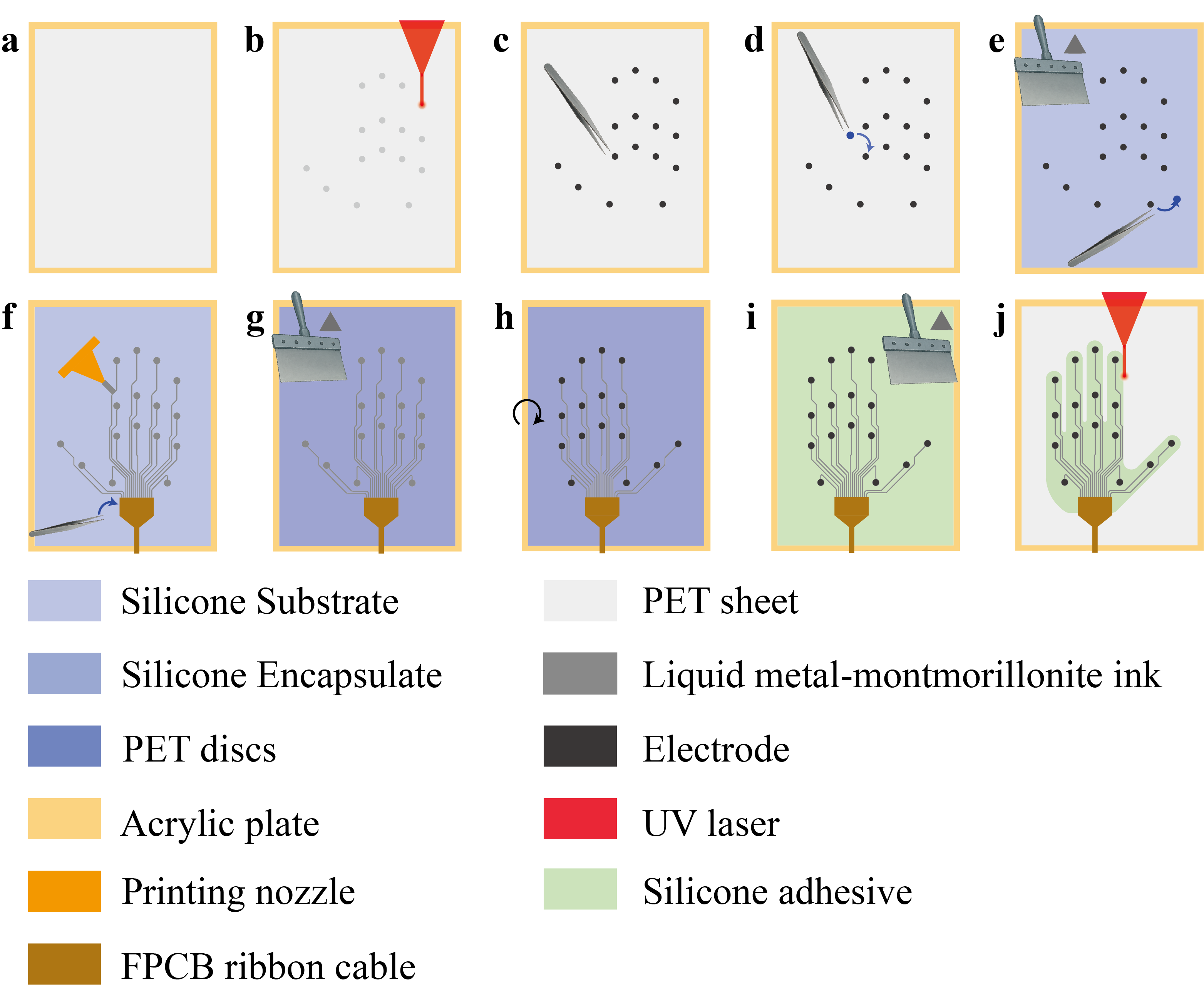}
  \caption{Fabrication of our electro-haptic device. (a) An acrylic plate is prepared as a rigid
  substrate, and a thin PET sheet is attached as the flexible base layer. (b) The predefined
  electrode positions are marked on the surface using a UV laser. (c) Sixteen electrodes are
  precisely positioned on the substrate. (d) Protective PET discs are placed over each electrode.
  (e) A silicone layer is applied to encapsulate the electrodes while aligning the FPCB ribbon
  cable. (f) Conductive liquid metal--montmorillonite (LMM) ink is dispensed through a pneumatic
  nozzle. (g) The entire device is encapsulated. (h) The device is detached from the substrate
  and flipped over, with the electrode side facing upward. (i) Silicone adhesive is applied to
  bond the device to the skin. (j) Finally, the device is cut into a hand-shaped form using a UV
  laser.}
  \label{fig:fabrication}
\end{figure}

Fig.~\ref{fig:fabrication} shows the fabrication steps of our electro-haptic device. A
polyethylene terephthalate (PET) sheet with a thickness of 0.1~mm was attached to an acrylic
plate to serve as the substrate for device fabrication, aiding subsequent processing steps.
A UV laser system was used at low power to
mark the corresponding electrode positions on this substrate. The electrode material was
prepared as described in the previous section, and 16 electrodes (15 stimulating electrodes
and one return electrode) were placed at their designated positions. Pre-cut 6~mm diameter PET
discs were placed on top of each electrode.
Subsequently, the Ecoflex 00-30 precursor was coated onto the substrate to form a 0.2~mm thick
film by a doctor blade. While the silicone rubber was partially cured, the PET discs on the
electrodes were removed to ensure the electrode material remained uncovered.

The device was then carefully placed on the platform of a pneumatic dispenser. The
pre-processed liquid metal-montmorillonite composite conductive ink was printed onto the device
according to the conductor trace pattern. After printing, the device was placed in an oven at
70~\textdegree{}C for 2 hours. This post-processing step activates the conductivity of the ink
and establishes effective electrical connections with the electrodes.

Prior to applying the encapsulating layer of silicone rubber, a flexible printed circuit board
(FPCB) ribbon cable was positioned at the interface area of the device, ensuring tight contact
with the activated conductive traces. Similar to the previous step, the silicone rubber
precursor was doctor-bladed to form a 0.2~mm thick film, completely covering the entire
electrode and trace region and encapsulating the upper end of the FPCB ribbon cable within the
device. Notably, pigments can be added to the silicone rubber as required to meet aesthetic
preferences. The silicone rubber was cured by letting it sit at room temperature for four
hours.

To better secure the device to the user's hand, a skin-safe silicone adhesive (Skin Tite,
Smooth-On) was selected as the adhesive layer. The device from the previous step was detached
from the substrate and flipped over, placing it with the electrodes facing upwards. Analogous
to earlier steps, pre-cut PET discs were placed on the electrodes, and a 0.1~mm thick layer of
the adhesive was doctor-bladed onto the device surface. The PET discs were then removed from
the electrodes, and the assembly was left to cure for 20 minutes. Finally, the device outline
was cut using a UV laser system according to the pre-designed file. The section covered by the
FPCB was manually cut using a cutting knife.

\begin{figure}[tb]
  \centering
  \includegraphics[width=\columnwidth,height=100mm,keepaspectratio]{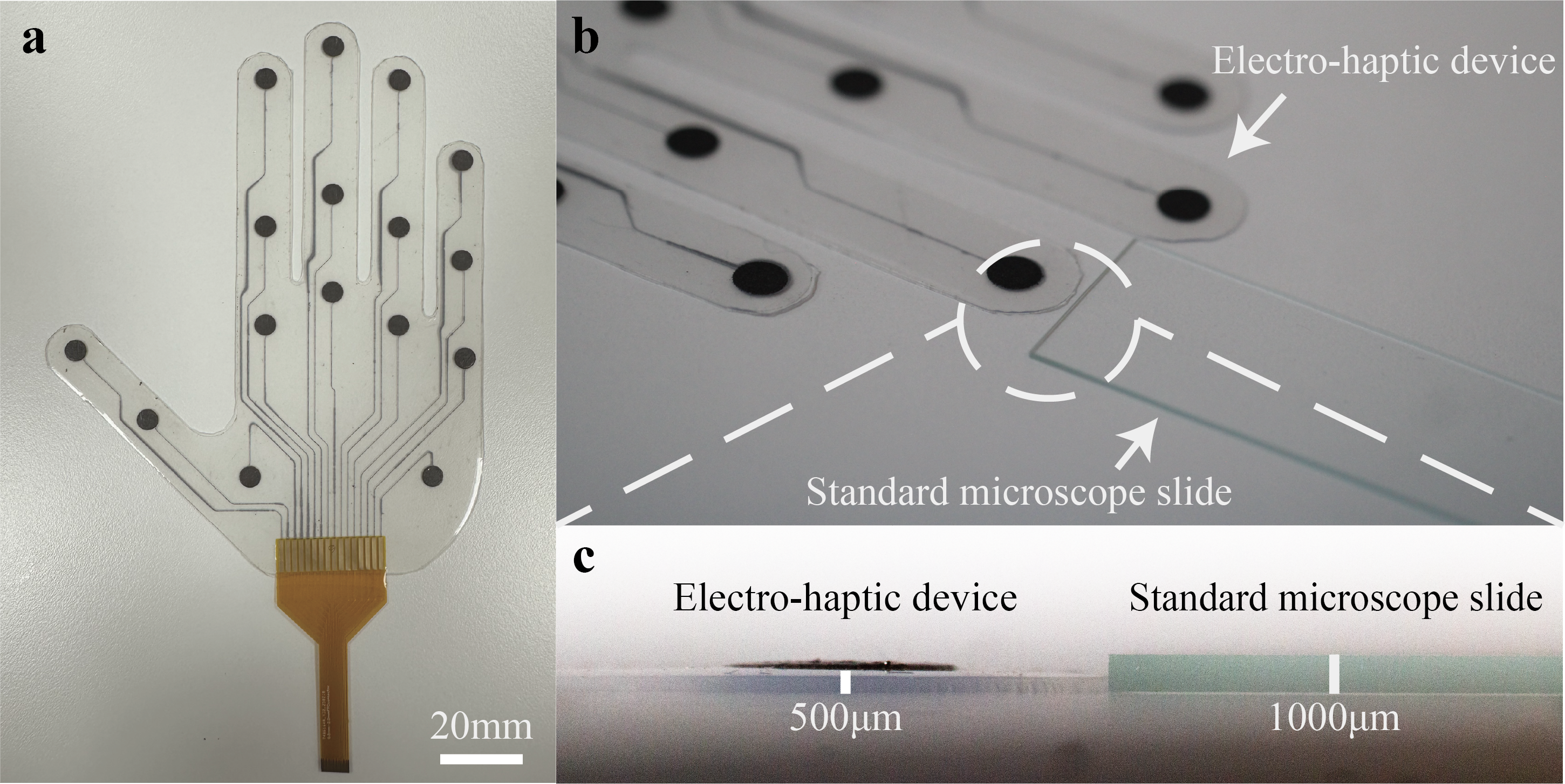}
  \caption{Fabricated electro-haptic device. (a) Front view of the device.
  (b) Close-up of the device beside a standard microscope slide. (c) Side view of the device.}
  \label{fig:device}
\end{figure}

The connection between the device and the custom drive system is achieved via the pre-installed
FPCB ribbon cable, accommodating wrist movement. To protect the circuit system from potential
failure due to factors like skin perspiration during use, it was encapsulated within a silicone
rubber mold, leaving only necessary electrical connection openings exposed. Similarly, using
the silicone adhesive, the circuit system can be affixed to the arm without additional
fixtures, thereby avoiding any sense of restraint on the arm. The picture of our device is
Fig.~\ref{fig:device}(a). The total weight of our device is only $\sim$48~g.

\subsection{VR Integration}

We developed our VR application with Unity3D for the Meta Quest 2 headset with real-time hand
tracking. We utilize Unity's physics engine to simulate realistic physical
behaviors---such as gravity, inertia---on virtual objects, enabling users to naturally grasp,
release, and manipulate items within the immersive environment. To further enhance realism, the
system dynamically adjusts object responses based on hand velocity and applied force, allowing
subtle interactions like sliding, pushing, or tossing to feel intuitive and physically
grounded. Additionally, we have implemented an improved collision detection algorithm between
the hands and virtual objects, ensuring that every contact point is captured with high spatial
precision. This refinement prevents visual artifacts such as object penetration and allows the
system to generate accurate symbolic representations of contact events at the software layer. These signals are then transmitted as serial input to the hardware interface,
enabling the connected stimulation module to respond with precise and low-latency feedback.

\section{Results}

\subsection{Performance of the Personalized Design Algorithm}

\begin{figure}[tb]
  \centering
  \includegraphics[width=\columnwidth,height=100mm,keepaspectratio]{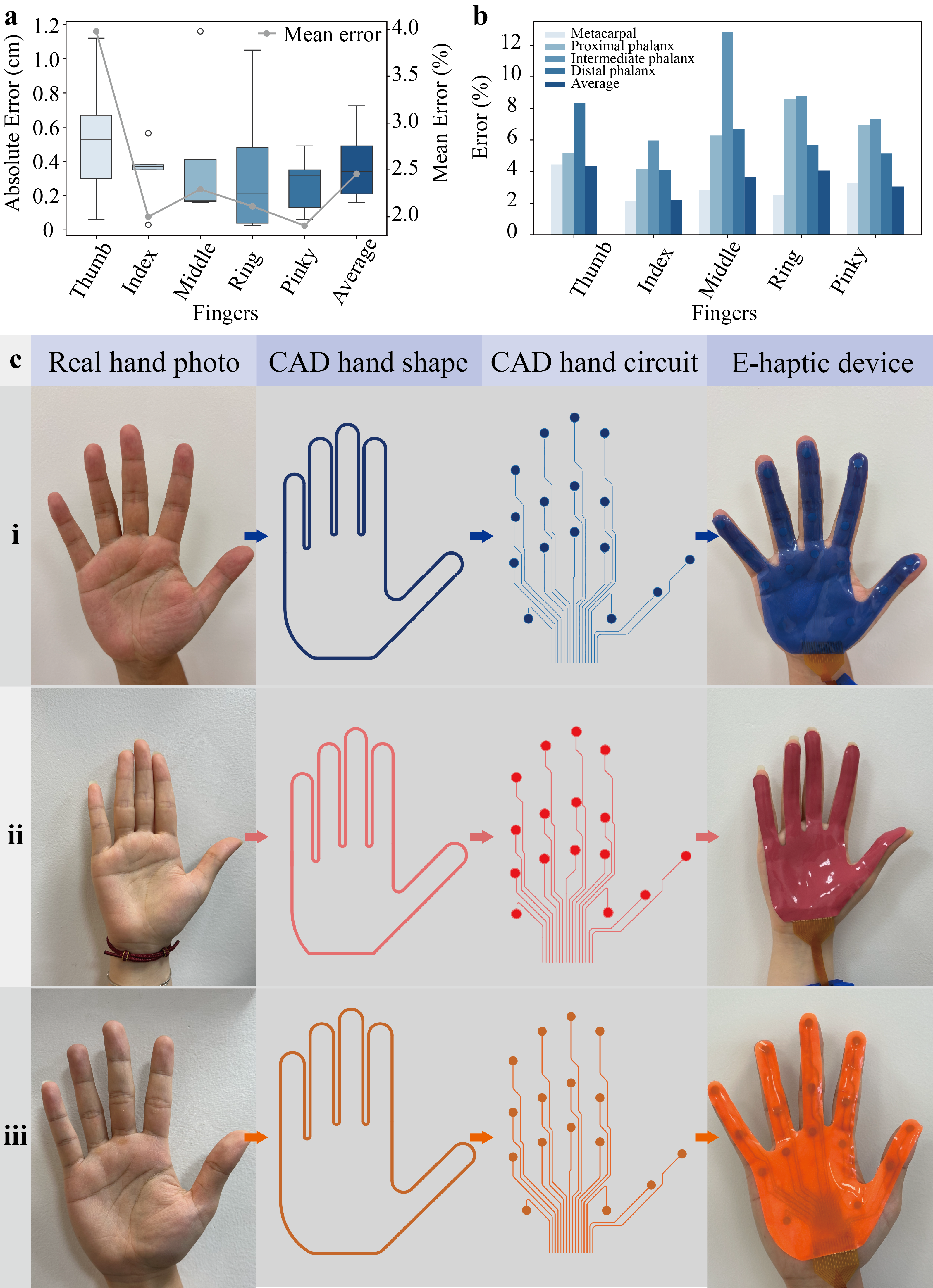}
  \caption{Accuracy and application of the personalized design algorithm.
  (a) Boxplots of absolute finger-length errors (cm) and a gray line showing the
  mean relative errors (\%) for each finger and the overall average.
  (b) Relative errors (\%) by anatomical segment (metacarpal, proximal phalanx,
  intermediate phalanx, and distal phalanx) and the average for each finger.
  (c) Personalized design workflow for three users (i)--(iii), showing, from left
  to right, the input hand photograph, CAD hand contour, electrode and interconnect
  layout, and fabricated electro-haptic device worn on the hand.}
  \label{fig:design-accuracy}
\end{figure}

To quantitatively assess the precision of the proposed personalized glove design algorithm, we
performed a statistical evaluation of finger length estimation on all five fingers, as
summarized in Fig.~\ref{fig:design-accuracy}(a) and (b). We collected data from the right hands of 10
volunteers (7 male and 3 female, aged 18--25, average age 21.70, $\mathrm{SD}=1.83$), all of
whom are right-handed, and compared the predicted finger lengths of our algorithm with
ground-truth measurements.

As shown in Fig.~\ref{fig:design-accuracy}(a), the absolute error (cm) between predicted and
ground-truth finger lengths is visualized using boxplots. The thumb exhibits the largest
variability, with absolute errors exceeding 1~cm in certain cases, while the forefinger and
pinky demonstrate relatively low error values, indicating more consistent predictions. Mean
absolute errors remain below 0.5~cm for most cases. The line shows the mean relative error for
each finger, with an average error rate below 2.5\% across the five fingers.
Fig.~\ref{fig:design-accuracy}(b) provides a
segment-level analysis by dividing each finger into four anatomical regions (metacarpal,
proximal phalanx, intermediate phalanx, and distal phalanx). We observe that intermediate
phalanx consistently yields the highest relative error, whereas metacarpal and distal phalanx
show lower error rates, highlighting the algorithm's better capability in capturing proximal
and distal structures.

\subsection{Application of the Personalized Design Algorithm}

The personalized design algorithm addresses the limitations of traditional
``one-size-fits-all'' gloves by using key hand parameters---finger length and joint
positions---to create a customized glove structure and functional layout. The algorithm begins
by constructing a parametric hand model from input anthropometric data. Finger length directly
determines the dimensions of the glove fingers, while the relative positions of the joints are
used to algorithmically determine the placement of functional components such as electrodes and
liquid metal conductive traces. Fig.~\ref{fig:design-accuracy}(c) demonstrates the application of
the algorithm to produce customized electrode layouts. We present three representative examples
that visually convey the pipeline---from capturing the raw hand image, to contour generation
and stimulation site mapping, and finally to the fabricated device applied on the hand. In
summary, this method enables the simultaneous customization of glove shape and functional
element placement, providing a scalable solution for producing truly personalized gloves that
are both comfortable and effective.

\subsection{Electrical Stimulation Waveforms}

\begin{figure}[tb]
  \centering
  \includegraphics[width=\columnwidth,height=100mm,keepaspectratio]{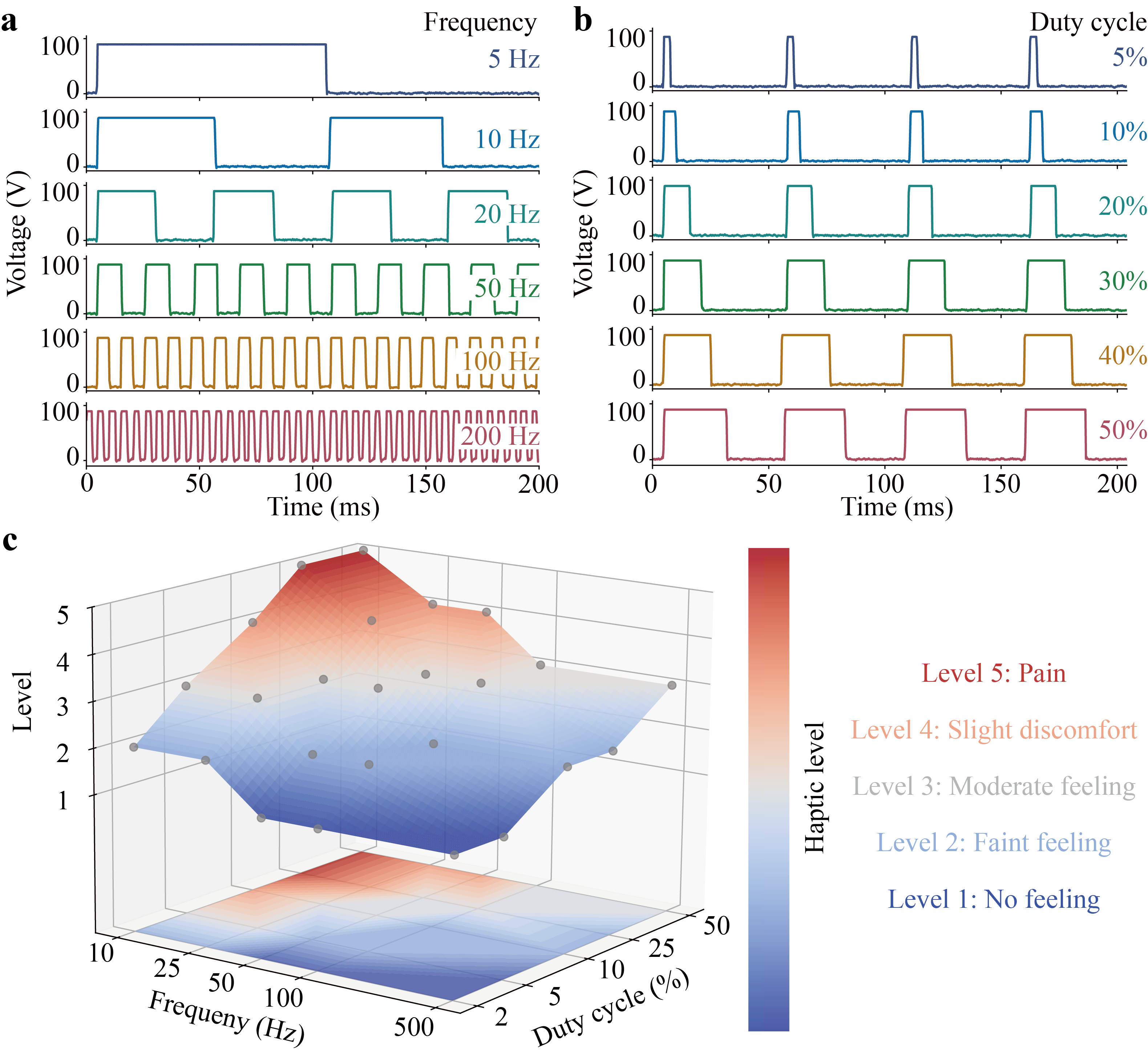}
  \caption{Electrical stimulation waveforms and perceived electro-haptic intensity.
  (a) Voltage waveforms at different frequencies. (b) Voltage waveforms at different
  duty cycles. (c) Mean subjective haptic ratings for the tested combinations of
  stimulation frequency and duty cycle on a five-level scale from no feeling
  (Level~1) to pain (Level~5).}
  \label{fig:waveforms}
\end{figure}

The stimulator delivers square-wave pulses with adjustable frequency and duty cycle. The
pulse settings used in this work included frequencies of 10, 25, 50, 100, and 500~Hz and duty
cycles of 2\%, 5\%, 10\%, 25\%, and 50\%. Fig.~\ref{fig:waveforms}(c) shows the mean subjective
haptic ratings for these parameter combinations. Fig.~\ref{fig:waveforms}(a) and (b) show stimulation
waveforms at different frequencies and duty cycles.

\subsection{VR Applications}

\subsubsection{Fine-Grained and Multi-Dimensional Electro-haptic Feedback}

VR provides a testbed for our device. In VR, users
often rely primarily on vision to infer the state of hand--object interactions, which can
increase task difficulty and diminish immersion. Most existing haptic devices
concentrate feedback at the fingertips or apply uniform, whole-palm stimulation due to limited
haptic spatial resolution and integration. Our electro-haptic device delivers multi-channel,
location-specific feedback across the hand, enabling more fine-grained tactile differentiation.

As illustrated in Fig.~\ref{fig:vr-applications}(a), users interact with different virtual
objects using different hand poses. Stimulation sites vary with contact location and gesture.
Users can also feel simulated airflow from a virtual fan on the palm. When the same object
is approached with different gestures (Fig.~\ref{fig:vr-applications}(b)), the stimulated regions
differ, producing distinct and realistic tactile qualities that enhance human--object
interactions.

\subsubsection{Haptic Timing Cues in VR Basketball}

Our electro-haptic device enables more precise and timely actions in VR, thereby increasing
realism and immersion. We developed a VR free-throw mini-game
(Fig.~\ref{fig:vr-applications}(c)) to evaluate how our electro-haptic device supports shooting
performance. In this task, participants performed stationary shooting without dribbling---they
simply picked up a virtual basketball and attempted to score. During each shot attempt,
participants used their right hand to grasp the ball from below while receiving sustained,
low-amplitude electro-haptic feedback indicating maintained hand-ball contact. Upon ball
release, the tactile stimulation immediately ceased with a brief, distinct pulse marking the
moment of hand-ball separation. The virtual ball then followed realistic physics, traveling
toward the basket with appropriate momentum and trajectory based on the player's throwing
motion. After each shot attempt, regardless of whether the ball scored or missed, participants
pressed a button to spawn a new basketball and reset for the next attempt. The system
automatically tracked shooting performance, recording both successful scores and remaining ball
count throughout each session. This design enabled players to focus entirely on shooting
mechanics and aiming, using electro-haptic feedback to precisely time their grip and release
without requiring visual attention to hand-ball interaction. The event-contingent electrical
cues supported critical shooting parameters including grip stability and release timing,
allowing users to keep their eyes on the target.

\subsubsection{Haptic Augmentation for VR Painting}

\begin{figure}[tb]
  \centering
  \includegraphics[width=\columnwidth,height=100mm,keepaspectratio]{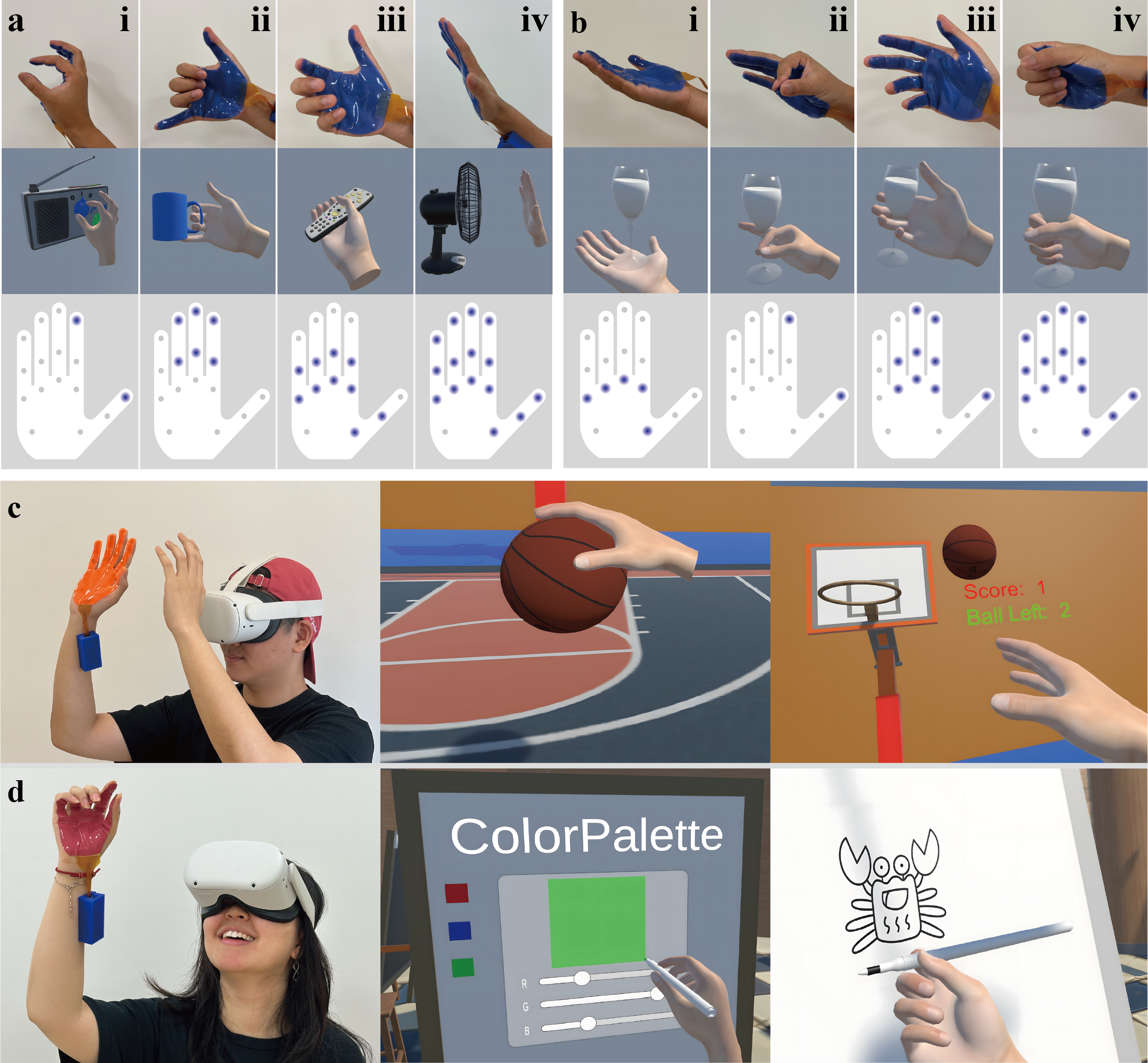}
  \caption{Electro-haptic feedback in VR applications. (a) Interactions with different
  virtual objects, showing (i) adjustment of a radio knob, (ii) holding a mug,
  (iii) holding a remote control, and (iv) feeling airflow from a fan.
  (b) Four gestures with a virtual wine glass, showing (i) palm-up holding,
  (ii) finger pinch, (iii) open-palm grasp, and (iv) full-hand grip. In (a) and
  (b), the rows show the real hand poses, corresponding VR interactions, and
  stimulation sites from top to bottom. (c) VR basketball mini-game with ball
  grasping and shooting. (d) VR painting with color selection and drawing on a
  virtual canvas.}
  \label{fig:vr-applications}
\end{figure}

Our electro-haptic device could further function as an effective tool for haptic augmentation,
broadening the range of tactile experiences. We implemented a VR painting environment
(Fig.~\ref{fig:vr-applications}(d)) to explore how electro-haptic feedback enhances creative
digital tasks. The virtual workspace provided three primary tools: a pen for drawing and
writing, a fill tool for area coloring, and an eraser for content removal. When participants
grasped the virtual pen, targeted electro-haptic stimulation was delivered to simulate
realistic pen-holding posture. This tactile feedback guided users toward proper grip
positioning without visual cues, supporting natural writing and drawing motions. The fill tool
and eraser provided similar grip-specific feedback patterns to differentiate tool selection
through touch alone. Color selection was facilitated through an RGB color palette interface
where users could mix primary colors to customize both pen and fill tool outputs. The palette
responded to touch interactions, allowing intuitive color sampling and tool customization
during the creative process. To enhance the sense of physical interaction, all tools
incorporated velocity-based force simulation. As users moved tools across the virtual canvas,
the system generated speed-dependent resistance that mimicked real-world drawing
dynamics---lighter, faster strokes felt smoother while deliberate, slower movements provided
increased tactile resistance. This velocity-responsive feedback helped users develop natural
drawing rhythms and pressure control. The painting surface consisted of a virtual whiteboard
where participants could freely draw, write, and erase content. The combination of
grip-specific tool feedback, color mixing capabilities, and velocity-based force simulation
created a multimodal creative environment that supported both precision tasks like writing and
expressive activities like artistic drawing, all while reducing dependence on visual tool
monitoring.

\section{Future Work}

The focus of this paper has been on developing a highly integrated and flexible user-customized
electro-haptic device capable of delivering precise and multi-dimensional feedback. While the
current study demonstrates the feasibility and effectiveness of our approach, there is room for
follow-up research and exploration.

\textit{Multi-modality feedback integration.} Our current device primarily provides
electro-haptic feedback. Future work could integrate additional modalities such as thermal
feedback (temperature cues) and mechanical feedback (e.g., pressure or vibration). Combining
these modalities would enable richer and more immersive haptic experiences, further enhancing
realism in virtual and augmented environments.

\textit{Sensor integration for enhanced interaction.} Incorporating integrated sensors (ee.g.,
motion tracking, force sensing, or bio-signals such as EMG/EEG) into the device would allow for
more precise detection of hand position, gestures, and fine motor control. Such functionality
could be particularly useful when visual tracking is limited---for example, in low-light
conditions or when the user's hands are outside the headset camera's field of view, supporting accurate
reconstruction of hand interactions in virtual spaces.

\textit{Applications in AR and XR.} While the present work focuses on VR scenarios, extending
the device to augmented reality (AR) and extended reality (XR) applications would broaden its
potential impact. In AR, for example, the device could support tactile cues for interacting
with virtual overlays on real-world objects, while in XR it could contribute to hybrid
workspaces, medical simulations, or remote collaboration environments where precise tactile
cues are important.

\section{Conclusion}

We introduced an untethered, ultra-thin, and user-customized electro-haptic device that advances
immersive human--machine interaction. Fabricated with soft silicone substrates and printable
liquid metal-montmorillonite ink, the device conforms to the hand while remaining lightweight
and minimally interfering. A palm-wide, addressable electrode array combined with waveform
modulation enables spatially precise, fine-grained, and multi-dimensional haptic feedback.

The user studies indicated minimal interference with hand movement, comfort during extended
wear, tunable tactile sensations assessed with established psychophysical methods~\cite{ref38},
and benefits in most VR scenarios, including improved overall shooting accuracy in the
basketball task.
We believe that our electro-haptic device can be applied to a broader range of human--computer
interaction scenarios and contribute to the advancement of next-generation haptic technologies.

\end{document}